# A Roman Dodecahedron for measuring distance

## Amelia Carolina Sparavigna

Department of Applied Science and Technology
Politecnico di Torino, C.so Duca degli Abruzzi 24, Torino, Italy

Here I am discussing a possible use of a Roman Dodecahedron, a bronze artifact of gallo-roman origin, for measuring distance. A dodecahedron, found at Jublains, the ancient Nouiodunum, dating from the 2nd or 3rd century AD, is used to create a model. Looking through the model, it is possible to test it for measurements of distance based on similar triangles.

The study of the surveying instruments of the ancient world is an interesting multidisciplinary research field, where experimental physics can be fundamental to appreciate the scientific and metrological knowledge of the ancient world. In fact, the study of some ancient objects can reveal their use as measurement tools, whereas archaeologists consider them as odd artifacts. For instance, I have discussed in some papers [1-3] the use of an ancient Egyptian balance case, found in the tomb of architect Kha, as an inclinometer suitable to measure the stair angles in surveying the building of subterranean tombs.

Recently, I became aware that there is a class of objects, the Roman dodecahedra, which are considered as a mystery of archaeology [4-6]. Fig.1 shows one of them. These are bronze artifacts of gallo-roman origin, having the dodecahedral form, dating from the 2nd or 3rd century AD. These objects exist in a variety of designs and sizes, always consisting of 12 regular pentagons. Roman dodecahedra have a diameter ranging from 4 to 11 centimetres. Some of them have at the center of the faces, holes of different sizes. Each of the 20 vertices is surmounted by one or three knobs, may be, to fit them on some surfaces. The Roman dodecahedra, about one hundred collected in several European museums, came from Gaul and the lands of the Celts: Great Britain, Belgium, Holland, Germany, Switzerland, Austria and Eastern Europe. According to several references, see for instance Ref.5, their function or use is considered as a mystery; because no mention of them has been found in the ancient literature.

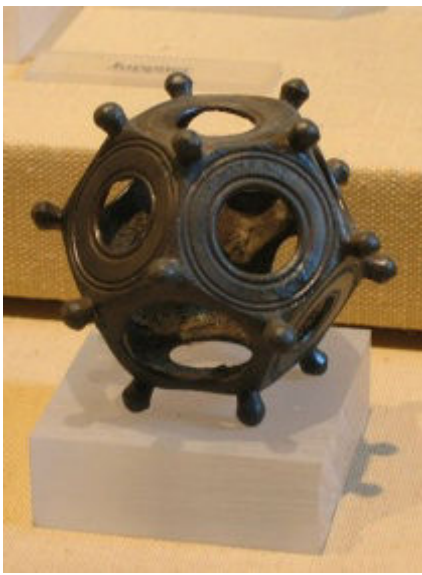

Fig.1 A Roman Dodecahedron (Source: Wikipedia)

There are so many theories reported by the World Wide Web about their use (candlestick holders; dice; survey instruments), often without a proper reference, that it is impossible to verify them. At a first glance of some of these dodecahedra, I guessed they were bowls for a sort of ancient bowling play [7]. But some of them look so complex, having holes of different size, that this hypothesis is weak. The hypothesis of the use of them as dice is weak too, because roman dodecahedral dice were quite different [8].

In 2010, a very interesting paper [4] reported of a new theory, proposed by Sjra Wagemans, DSM Research, assigning an astronomical feature to these objects. Wagemans used a bronze copy of a dodecahedron to see whether it was possible to determine the equinoxes of spring and autumn. As written in [4], "the dodecahedron is therefore linked to the agricultural cycle, both sophisticated and simple at the same time, to determine without a calendar, the most suitable period of time during the autumn for sowing wheat." Crops were of vital importance for the Roman legions located in regions far from Rome. What is remarkable is that Wagemans is using an experimental approach testing its device on a period of time of several years at different locations and latitudes [9].

Following an experimental approach too, I have prepared a dodecahedron, made of paper, according to the data of one of them given in Ref.10. In this reference we can find a detailed description reporting the sizes of the holes (see Fig.2). The dodecahedron, found at Jublains, the ancient Nouiodunum, is dating from the $2^{nd}$ or 3rd century AD. Here I am proposing this Roman dodecahedron for measuring distance using similar triangles. The dodecahedron possesses five angles of view, and, knowing the size of a distant object allows to determine its distance. Let us remember that in photography the angle of view is the cone which describes the angular extent of a given scene that is imaged by a camera. The dodecahedron can be used to have some specific angles of view.

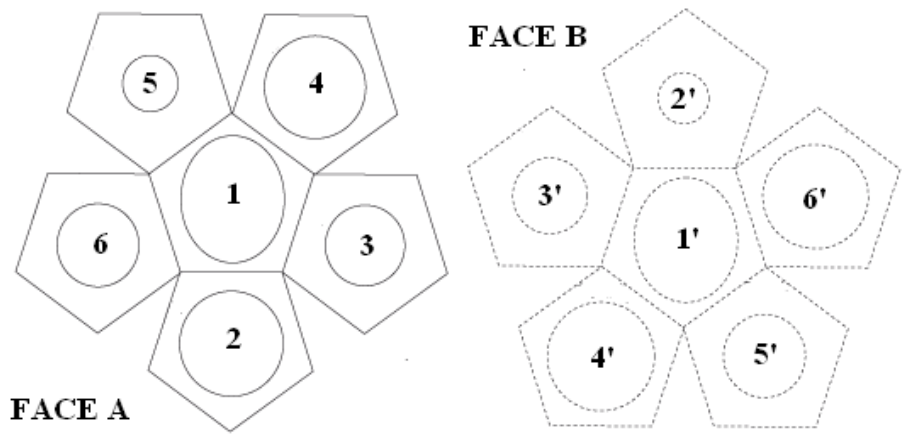

**Fig.2 The faces of the Roman Dodecahedron of Ref.10.**

Naming 1,2,3,4,5, and 6, the holes of face A, and 1',2',3',4',5' and 6' the holes, of face B (see Fig.2), we have the following opposite pairs: (1',1),(2',6), (3',5), (5',3), (4',4) and (6',2). The sizes of the hole are: 26×21.5 mm (1), 21.5 mm (2), 16.5 mm (3), 21 mm (4), 11.5 mm (5), 17mm (6), 25.5×21.5 mm (1'), 10.5mm (2'), 15.5 mm (3'), 22 mm (4'), 17 mm (5') and 22 mm (6').

Let us consider for instance, the pair (2',6) and look through the dodecahedron, holding it with 2' and 6 parallel, with 2' near an eye and 6 as opposite. If the dodecahedron is close the eye, we see the two holes; if it is too far, we see just the nearest hole 2'. There is a distance where we can see the circumferences of the two holes as perfectly superimposed. This is a specific angle of view that we can use for measurements (see Fig.3). We can decide this as the "correct" angle of view to properly use the device.

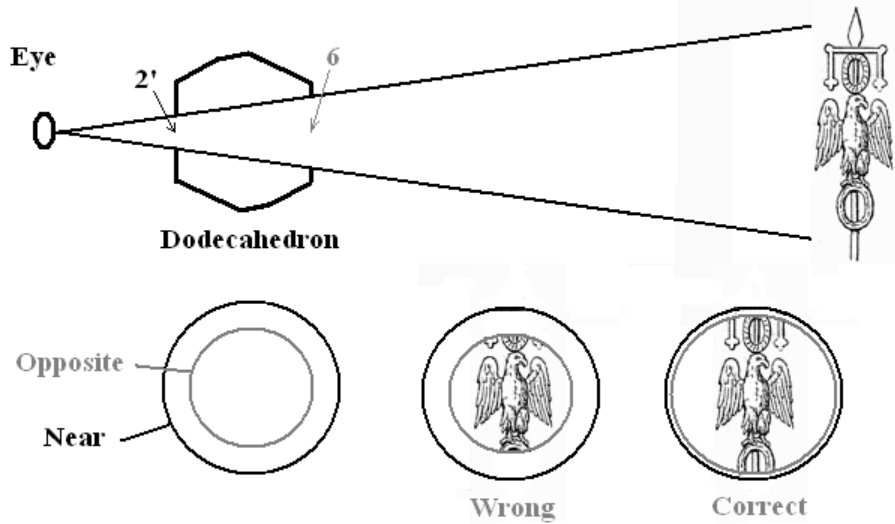

**Fig.3** Let us consider the pair (2', 6) and look at a Roman vexillum through the dodecahedron, holding it with 2' and 6 parallel, with 2' near an eye and 6 opposite. If the dodecahedron is close the eye, we see the two holes. If it is too far, we can see just hole 2'. There is a distance where we can see the circumferences of the two holes (black and grey in the image) as perfectly superimposed. This is a specific angle of view that we can use for measurements.

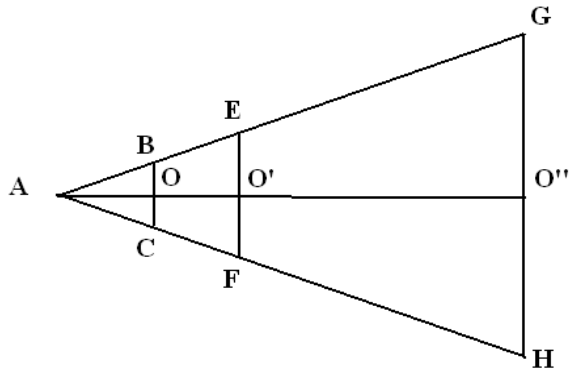

**Fig.4** Geometry of Fig.3

To shows how we can determine the distances, let us consider the following example. Let us imagine that a Roman soldier is looking through the dodecahedron at a vexillum, of which he knows the size, for instance, having a height of two meters. We could ask ourselves, if the soldier sees the vexillum fitted in the field of view of the pair (2',6), that is, with its height coincident with the diameter of the superimposed circumferences 2' and 6, what is its distance?

To answer, let us look at the similar triangles of Fig.4, where A is the eye of the soldier, BC the diameter of 2', EF the diameter of 6, and OO' the distance of the two holes. GH is the vexillum. From the figure, we can obtain that:

$$(EO'-BO)/OO'=(GO''-BO)/OO''$$

Let us suppose GO''>>BO, then (EO'−BO)/OO'=GO''/OO'', and therefore:

$$(EF-BC)/OO' = GH/OO''$$

We have;

$$OO'' = GH \times OO'/(EF-BC)$$

In the case of our soldier, who sees the vexillum fitted as GH, we have: $OO'' = 2\times10^3$ mm $\times$ 50 mm /(6.5 mm) $\approx$ 15 m, where we have assumed that the distance between the opposite faces, OO', of the dodecahedron is of 50 mm.

This is the result for the pair (2',6). Let us see what happens if he use the other pairs. The results are displayed in the following table.

| Pair | GO''/OO'' |
|---|---|
| (2',6) | 0.065 |
| (3',5) | 0.04 |
| (4',4) | 0.01 |
| (5',3) or (6',2) | 0.005 |

Using these data, for the vexillum of 2m, we have distances of 15 m, 25 m, 100 m and 200 m, respectively. Then the soldier can used the device for four different ranges. Besides them, there is also the pair (1,1'), that can be used.

It seems therefore, that looking through the dodecahedron as previously discussed at an object, the size of which we are able to estimate, we can have an approximate distance of it. As previously told, Roman dodecahedra have been proposed for surveying and military purposes. Unfortunately I was not able to find any reference about these studies. After preparing a simple specimen, my conclusion is that the hypothesis of a use for surveying is stronger than that previously proposed of these artifacts to be dice or bowls. Moreover Ref.10 is telling that the dodecahedron was found in the same place where a balance was found, the house of people selling precious stone. May be, besides precious stones, the seller offered also tools for measurements such as balance and dodecahedrons. However, authors of Ref.10 are supporting the hypothesis of a divination die.

Since the sizes of dodecahedrons are quite different, further studies are necessary to test different models.